# Impact of Covid-19 on Taxi Industry and Travel Behavior: A Case Study on Chicago, IL

Naga Sireesha Chinthala[1], Jenell Lewis[1], Sravan Vuppalapati[1], Khiran Kumar Chidambaram Sivaraman[1], Chinmay Vivek Toley[1], and Huthaifa Ashqar[1, 2, *]

[1] Department of Computer Science and Electrical Engineering, University of Maryland Baltimore County (UMBC)
[2] Arab American University, Jenin, Palestine
[*] Corresponding Author

## Abstract

As the debate over the future of transportation continues in the midst of the COVID-19 pandemic as a deepening global crisis, taxi industry seems to be not spared by the quick and disrupting changes that may arise from the pandemic. The impact is relatively higher in major cities because of the high-density population and transportation congestion. In this study, we used spatial analysis and visualization to investigate the impact of the pandemic on the economics of the taxi industry and travel behavior using trip-by-trip data from the year of 2014 to 2020 in Chicago, IL. Results show that there is a drastic decline in the trips in the central city and airport areas. During the pandemic, people tended to travel longer distances, but travel times were considerably less because of the significant reduction in traffic volumes. Results also showed that the top twenty most popular pick-up and drop-off locations only included Chicago Downtown and OHare International Airport before the pandemic. However, during the pandemic, the top twenty most popular pick-up and drop-off locations is distributed between the Airport, the Downtown, as well as many other areas along Chicago Eastside.

**Introduction**

WHO has declared COVID-19 a pandemic on 12 March 2020. The impact the pandemic has left on the economy is significantly high. Virus containment policies and measures such as social distancing, travel restriction, community lockdowns, work from home options, mandatory quarantines, restriction on public gatherings impacted people's travel behavior and as a result, the taxi industry has seen its worst year in 2020 in terms of revenue. The impact could last for longer than one has imagined and requires transformational changes in the travel industry. In major cities like Chicago and New York where the taxi industry is so prominent and is the main source of income, many families are hit extremely hard by the pandemic. In Chicago, most of the industries have started to recover with the release of vaccination and more events being organized. But it is not the same for the Chicago taxi industry, which was hit even before COVID, because of rideshare companies like Uber, Lyft operating in cities.

"The pandemic really fast-forwarded the decline of the taxi industry," said Mohammed. Nobody imagined a drop of more than 80%, that too for a duration of more than a year. "Garcia, who lost 85% of his business when the pandemic hit, said he's happy if he makes $1,000 in a week now, but has to pay insurance costs 800 a month" (Chris., 2021). The drivers and their families depending on income from riding taxis do not know how to deal with the pandemic. According to the city's official data portal, the active medallions number is only 800 but the actual number of taxis on the streets is far less than that (Chris., 2021). In this paper, we will see the impact based on different features like the number of rides, active taxis, revenues for the companies and community areas. We hope this paper will provide a bigger picture of the COVID crisis in the taxi industry and help in developing financial support and mitigation strategies.

**Literature Review**

*Spillover of COVID-19: impact on the Global Economy*

In the paper, the impact of Covid-19 on the global economy is analyzed[5]. The analysis is based on how the social distancing guidelines affected things, and how the uncertainty led to increased spending and investing. The analysis is done on the period from early 2020 through March 2020 when the pandemic began. The conclusion from the 30-day social distancing policy was that there was a negative impact on the economy which was caused by a lower amount of economic activity and a fall in stock prices. They also found that the pandemic had a positive effect on some healthcare systems by forcing them to fix their systems. This also applied to other systems in the public sector such as transportation and disease detection. Their analysis was good, but due to their limited dataset, they could not create a clear picture of the socioeconomic consequences of the government's policies.

*The fall and rise of the taxi industry in the COVID-19 pandemic: a case study*

The impact of Covid-19 on China's taxi industry was analyzed in this paper[6]. This analysis was based on a four-week period which encompassed both during and after Covid-19 in Shenzhen, China. They did a spatiotemporal analysis on their data and it was found that there was a decrease in taxi usage of almost 90%. Even after the city opened up, the usage did not return to the level it was before Covid-19 because more people opted to use their own cars. They also found that the city center had a larger reduction in trips compared to the suburbs.

*Impact of COVID-19 on Urban Mobility during Post-Epidemic Period in Megacities: From the Perspectives of Taxi Travel and Social Vitality*

This paper[2] concentrates on the impact the pandemic and associated prevention and containment strategies have on the taxi industry considering the major cities in China. The author has mentioned and analyzed different features that impacted the industry. The research was based on the taxi trip data from May 2019 and Jan - Jun 2019. Impact analysis is performed based on the number of trips, revenues respective to both spatial and temporal analysis. This paper has helped in shaping our project and provided insights on the course of action.

*Predicting the Taxi Fare Of Chicago Cabs*

This paper[7] is concentrated on predicting the fare of a taxi in Chicago for a given day of the week and time of the day. Though the research questions are not directly related, this paper has provided great insight into the data set and different possible issues with the data and taxi industry of Chicago. The background information regarding the prices, dataset, and Chicago community areas has a very great influence on our project.

*Assessing the Impact of Reduced Travel on Exportation Dynamics of Novel Coronavirus Infection (COVID-19)*

This paper[8] concentrates on the impact of the drastic reduction in the travel volume within China in January and February 2020 with respect to the COVID-19 cases outside China. They made use of statistical models to estimate the impact of travel reductions on three epidemiological measures. In their work, they have concluded that the impact of travel volume to and from China during the COVID -19, estimating that the time delay to the major epidemic was on the order of 2 days.

**Research Question**

This research is done in order to understand what impact the pandemic and associated strategies have on the taxi industry of Chicago. We will analyze the impact respective to different POIs (Point of Interest) which is the number of trips, number of active taxis, companies, trip distances, and locations.

**Materials and Methods**

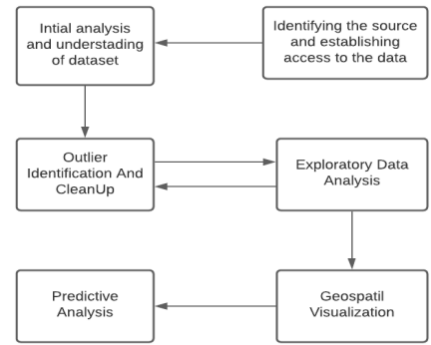

The City of Chicago has been collecting trip-by-trip taxi data from 2013, and the data has been made publicly available on Chicago's open data portal (https://data.cityofchicago.org/). The collective data from 2013 till present is made available on google cloud          Figure 1: Research Framework as part of public datasets. We have created a Google cloud service account and JSON with service credentials to access the dataset. The initial analysis is performed to understand the trend and features of the dataset. Exploratory analysis is performed to produce the results respective to each feature of interest. Using GeoPandas and Folium (libraries in python used for geospatial visualization) compared the impact on community areas. Random Forest and Decision Tree models are used for predictive analysis, to understand the trend and impact.

*Dataset*

The dataset which is available as part of the public data set with the name "chicago_taxi_trips" is accessed using a google cloud service account. The total number of rows in the dataset is 187 Million. Dataset can be accessed using the below URL.

"https://console.cloud.google.com/bigquery?p=bigquery-public-data&d=chicago_taxi_trips".

Dataset consists of unique_key, taxi_id, trip_start_timestamp, trip_end_timestamp, trip_seconds, trip_miles, pickup_census_tract, dropoff_census_tract, pickup_community_area, dropoff_community_area, fare, tips, tolls, extras, trip_total, payment_type, company, pickup_latitude, pickup_longitude, pickup_location, dropoff_latitude, dropoff_longitude, dropoff_location columns. Taxi_id column is encrypted to ensure the safety of drivers, trip start, and end timestamp columns are rounded to the nearest 15 minutes, pickup and dropoff latitude and longitude columns are not the exact locations, they are accurate only to the census tract and community area level, census tract information is provided only if at least 3 trips are reported during the 15-minute duration. These masking measures and missing information are important to maintain the privacy of taxi customers. Given the nature of data reporting, the data cannot be used to track or analyze the trips which are currently taking place or just completed.

*Data Cleaning*

1. Trips with trip end time less than trip start time are found in the data set. Trip completion even before the start is not valid and this kind of data is removed from the final dataset for accurate results.

2. Both trip end dates and trip seconds are missing for few records. Since both variables are missing it is not possible to identify the duration and exact time of the trip. These records are removed from the final dataset.

3. Records with trip miles of more than 100 are eliminated from the final dataset. 100 miles is considered because the longest route within Chicago which is from O'Hare international airport to Hegewisch is around 44 miles, even considering round trips it should not be more than 100 miles, and any trips with more than 100 miles must have happened outside Chicago or reported wrongly.

4. Trips with a trip duration of more than 3 hours are also dropped from the final dataset. The reason being the longest trip from O'Hare to Hegewisch would only take 1 hour 30 minutes during peak hours, we are considering double the time reported in peak time.

5. As per Chicago's official regulation, the base fare for any trip will be $3.25. So any trips reported less than $3.25 are removed.

6. According to the dataset description on Chicago's data portal community area information is null for the trips reported outside Chicago. 18M records which is 20% of the data is missing area information, since our focus is to analyze trips that happened inside Chicago, these trips are dropped from the analysis.

7. Most of the fields for 2013 data are missing, and 2021 is incomplete and continuously changing, so the data for these years 2013 and 2021 is not considered in the analysis.

**Results and Discussion**

*Statistical Analysis*

Taxi industries performance has been on a steady decline from 2014 until 2020 March, before covid hit the United States, on March 12 with COVID being declared pandemic and restrictions on travel there is a drastic decline in the number of trips per day. The average number of trips per day which was around 50000 before the pandemic has come down to less than 5000. The graph from Figure 1 also shows seasonal behavior which is repeating for every year. Even at the end of the year 2020, the number of trips is still far below the pre-pandemic levels. Every year there is a rise in the number of trips in March except for the year 2020. 23000 trips were reported on March 1st, 2020, and only 2125 trips were reported on the last day of March, which is only 10% of the first day trips. The number of trips reported in March 2020 was 50% less than trips reported in February. April has got the worst of the pandemic and the number of trips further went down by 90% compared to March. Though it showed some improvement after April, the trip count is still far from recovery.

Figure 2: The number of Trips for each day from 2014-2020

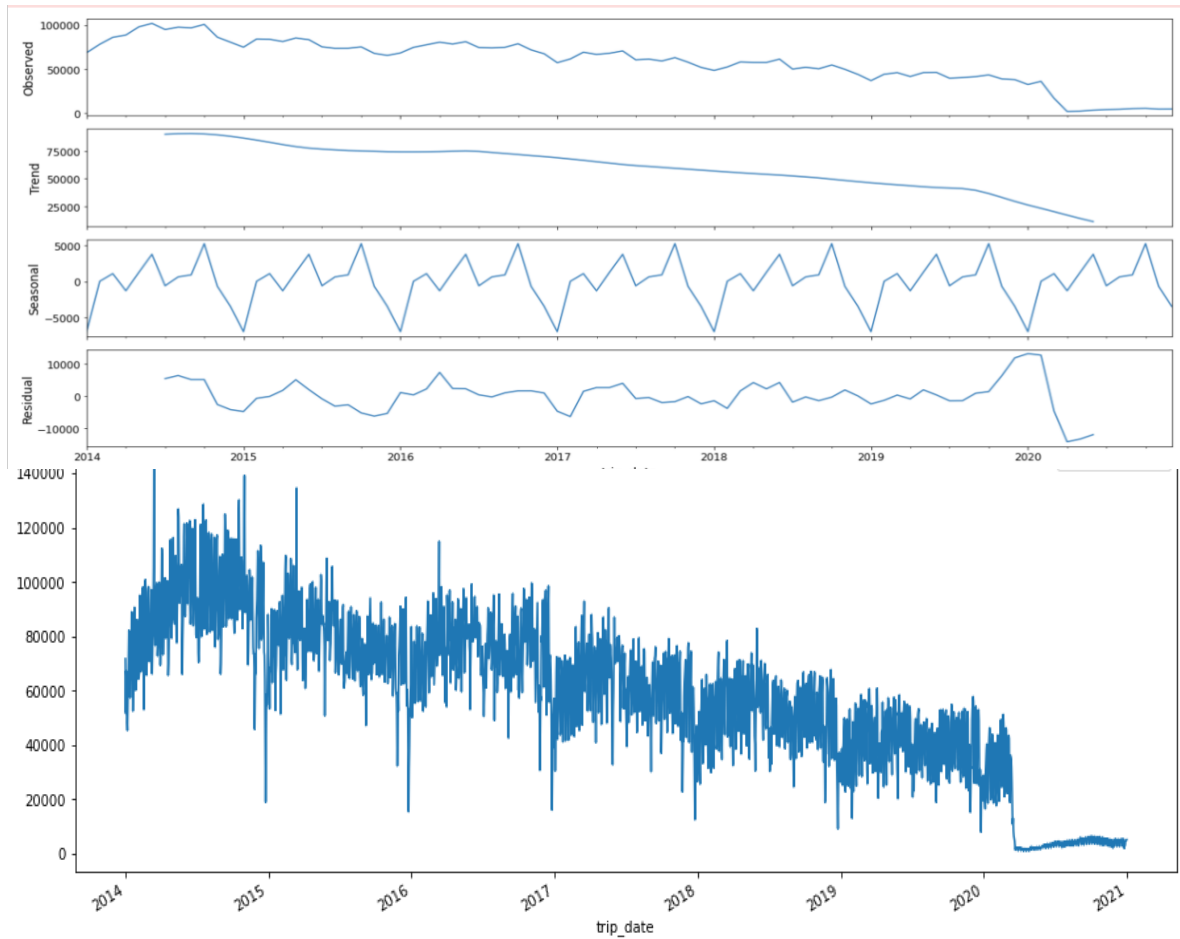

Figure 3: Statistical analysis of Trips from 2014-2020

*Average Trips Per Hour*

The number of trips that take place on weekdays is nearly double the trips that take place on weekends. The pandemic did not impact people's preferred times of travel. 4 pm to 6 pm remained peak hours for operation on both weekdays and weekends, with an average of 800 and 400 trips per hour respectively which were otherwise 1600 and 3200. Midnights of the weekend remained the preferred time of travel on weekends.

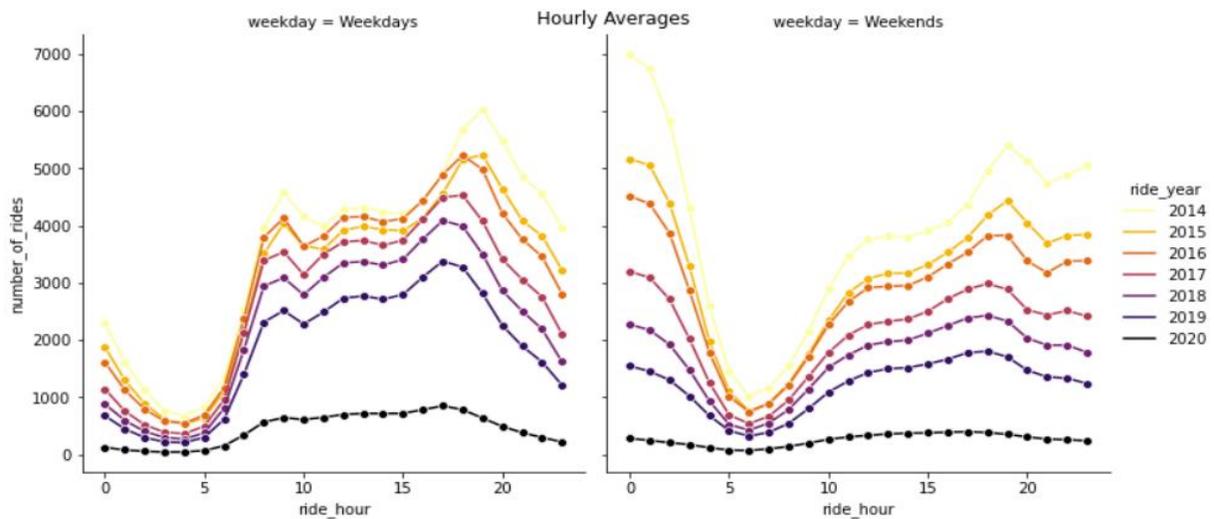

Figure 4: Average number of trips for each hour from 2014-2020.

*Impact on Taxi Companies*

The amount that is collected from trips for each company has gone down drastically, most of the companies suffered more than 80% losses. The only company which has increased in the amount made from trips is U Taxi Cab. It is observed that this company was operating with only 7 taxis before the pandemic hit, the number increased to 24 by the end of 2020.

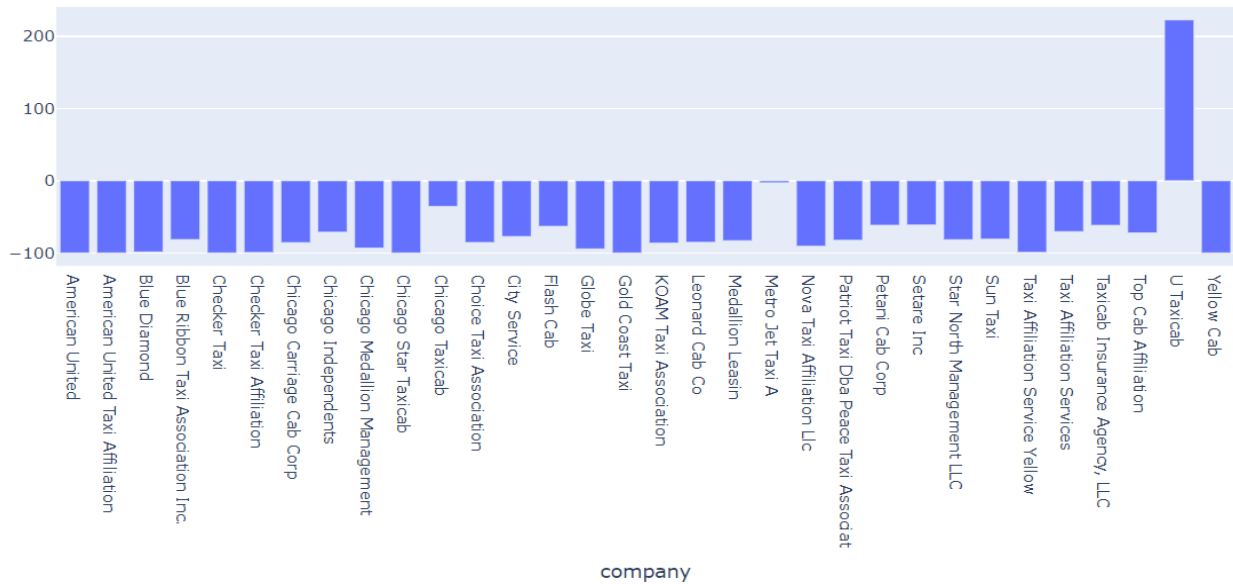

Figure 5: Gain and loss from trip amount per company for each year from 2020 vs 2019

*Active Taxis*

With the number of trips going down from year on year, the number of medallion licenses was also decreasing. According to the City of Chicago, there were nearly 6000 active licenses in the year 2018, and more than 8000 in the years 2014 to 2016. It was reported that only just more than 800 active licenses were present during the pandemic, but the actual number on the road was just around 400 in May.

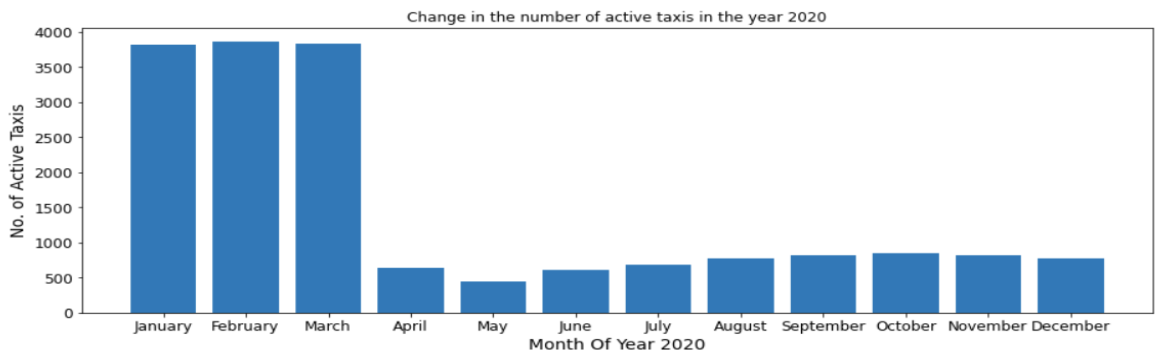

Figure 6: The Number of Active Taxis for each Month in 2020.

*Travel Times and Distances*

The pandemic has shown an impact on travel distances, especially on weekends. People tended to travel longer distances than usual, but travel times were considerably less because of a reduction in traffic until last quarter. Last quarter has seen a rise in both average travel distances and travel times.

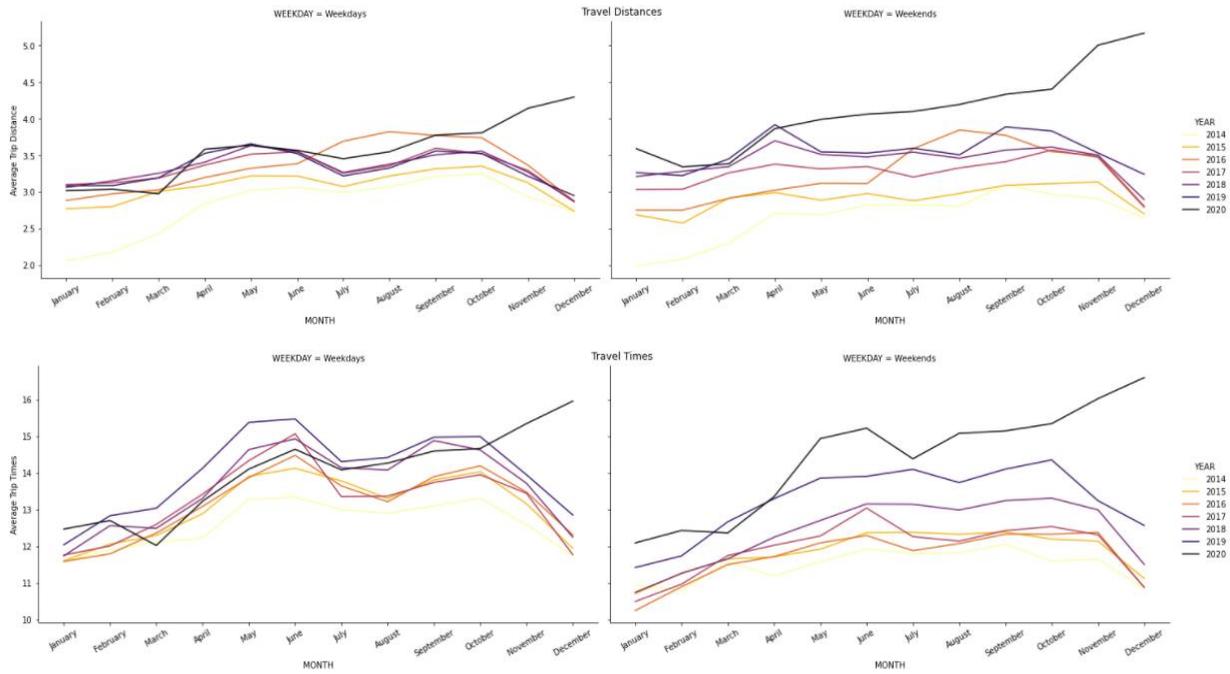

Figure 7: Comparison of the Travel Times and Distances

*Tip Behavior*

In contrast to the number of trips, the average tip amount was rising consistently from 2014 until 2020 March. Pandemic has hit even the generosity of people and the average tip amount fell below the average of 2014, adding additional financial burden on the drivers.

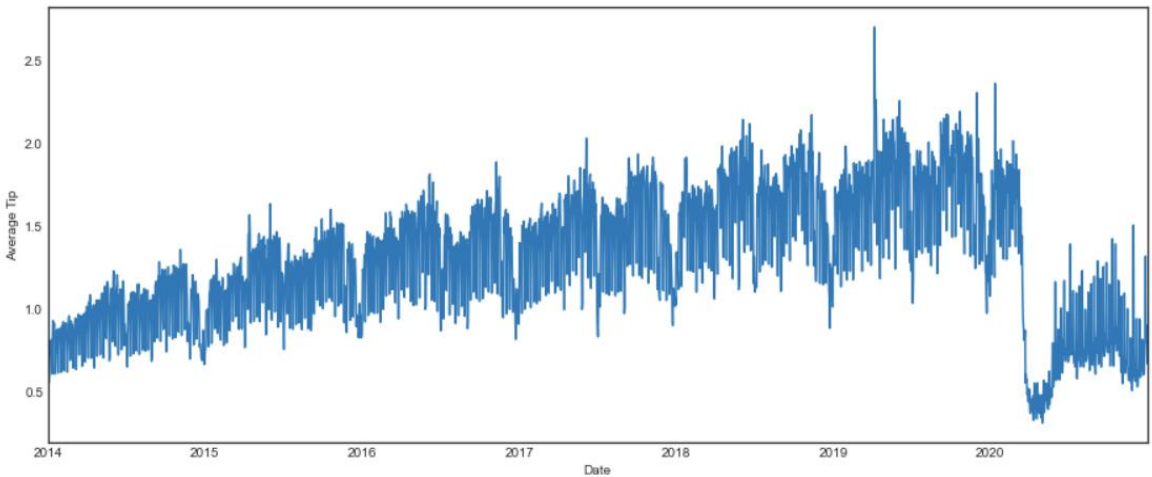

Figure 8: The trend of Tips throughout the years.

*Wait Times for Next Pick-ups*

With fewer people on the road and people traveling outside or coming into the state reducing drastically, the taxi drivers are spending hours in the cab waiting for the next pick-up. Taxi drivers are waiting hours to pick up a single fare at O'Hare Airport or outside Union Station (Lourdes., 2020). Most drivers get the net pick-up within 100 minutes of the drop-off earlier, but the average wait time has increased to 400 minutes. Some drivers said they're now sleeping in their cars at O'Hare so as not to lose their place in line (Lourdes., 2020).

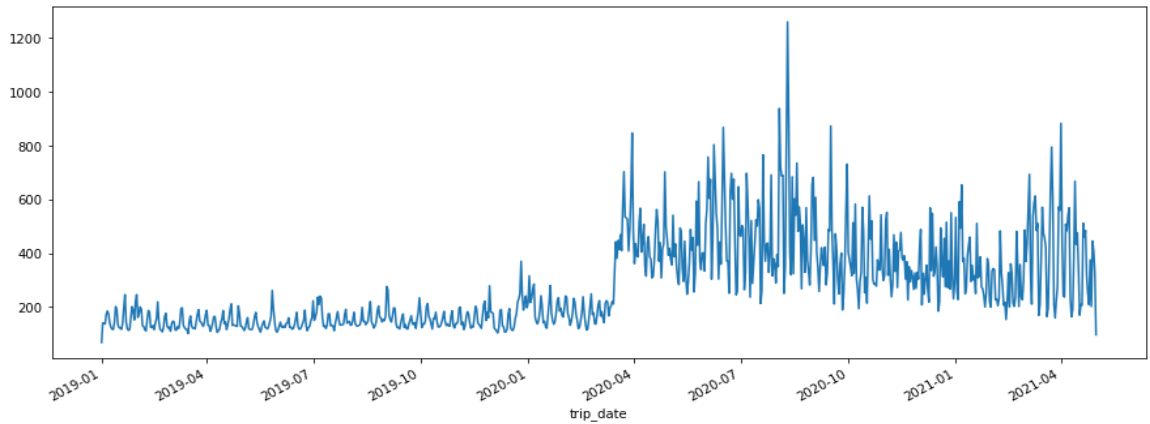

Figure 9: Average wait time for the next trips

*Number of Trips Per Taxi*

There were 12 trips per taxi on average until March-2020. As shown in Figure 10 the number of trips per taxi reduced to 6, that too with the limited number of taxis operating as we mentioned in the earlier section the number of active taxis has gone down from nearly 4000 to 800. It is difficult to say if the drivers did not hit the roads or they did not get any pickups. Considering they might have got at least one trip if they were on the road, if all taxis were operating then the number of trips per taxi should have gone down to 3 or even less.

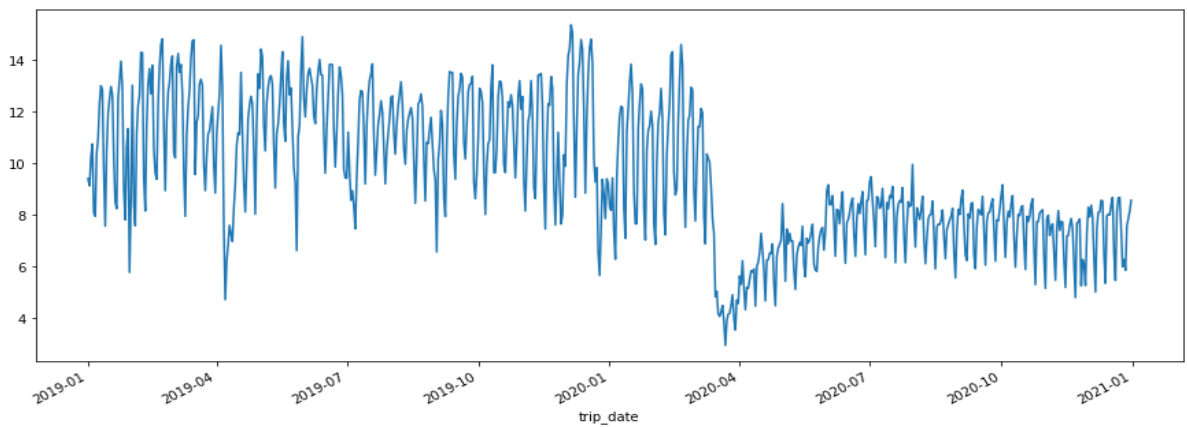

Figure 10: Average number of pick-ups per taxi

*Pick-up and Drop-off Respective to Community Areas*

The Loop and downtown are the major business districts of Chicago. These areas contribute to nearly 70% of the total taxi business that happens in the city. It is also the major area for the taxi industry, before the pandemic the average number of pickups and dropoffs was around 45000 per day, which went down to 2500 during the pandemic. O'Hare International Airport is the 4th popular area for taxi usage in the City, with an average of more than 3000 pick-ups and drop-offs per day, but with restrictions on multiple countries to enter the United States and flights being operated in limited capacity there were fewer than 300 trips on average in 2020. Garfield Ridge on the Southwest side is another popular area with more than 1000 trips per day before the pandemic, but the average number of trips during the pandemic is less than 100 trips. There is an overall 95% decrease in the number of trips in most of the popular areas. The impact is considerably less in the North and Southwest side of the city (Fig.12). As shown in Figure 11, the Far Southeast side is seeing an increase in the demand for past years, and even during the pandemic, there is a rise in the number of pickups in these areas in contrast to all other places.

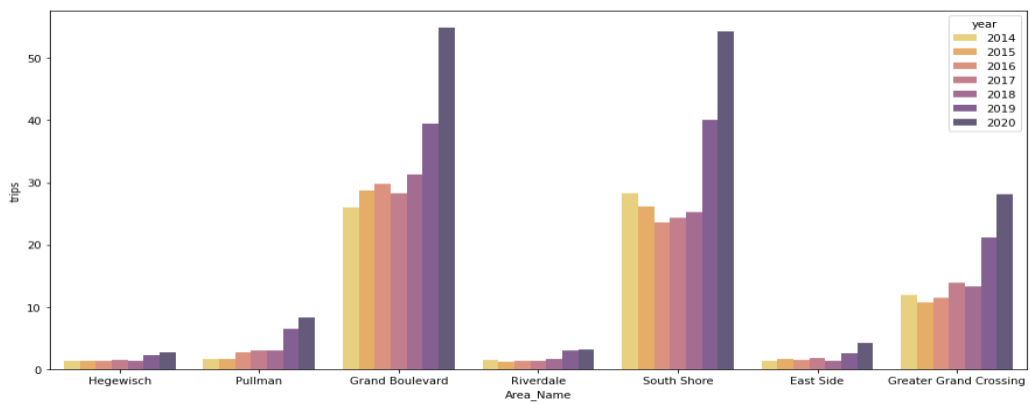

Figure 11: Trips in Far Southeast side of the city

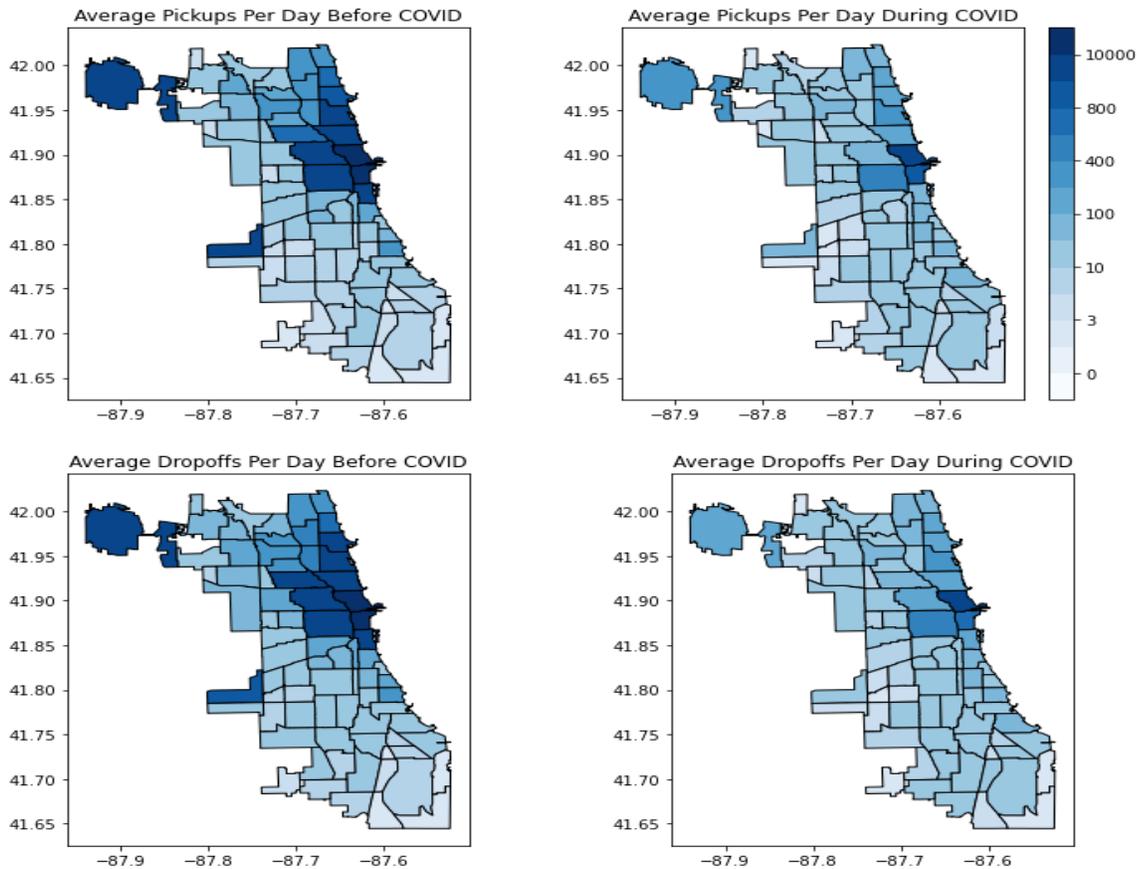

Figure 12: Density comparison of pick-ups and drop-offs

Few places have doubled the regular pick-ups and drop-offs compared to before the pandemic. Figure 13 shows these locations, green represents the pickups and red represents drop-offs. These places include Calumet Ave, South Throop St, Kennedy King college area, South Paulina St, West 47th St, and few others.

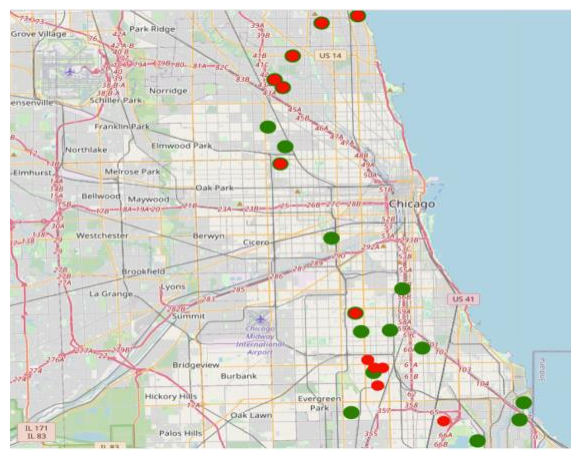

Figure 13: Doubled trips during the pandemic

The top 20 most popular pick-up and drop-off locations included only downtown and O'Hare international airport before the pandemic. It is observed that the top 20 spots during Covid also included the Airport, downtown, areas along the length of Chicago's Eastside, Central Chicago, and an area in West Chicago (Fig.14).

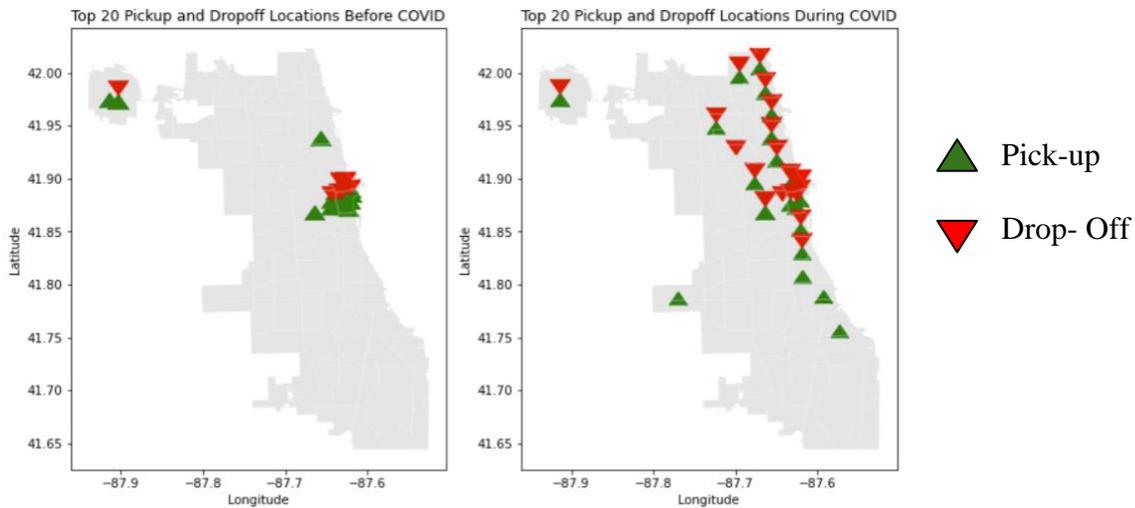

Figure 14: Top 20 pick-up and drop-off locations before and during COVID

*How 2020 For Taxi Industry Would Have Been Without Pandemic*

What the taxi industry business would look like without the pandemic is very interesting to know. To understand this, we have used Random Forest (RF) and Decision Tree (DT) algorithms to predict the number of trips for 2020. Though the trend looked somewhat linear, when we tried to fit the data for Linear Regression it did not give expected accuracy as the difference in the number of trips was relatively high, it failed to fit the linear model. New variables Year, Month, Hour and Weekday are created for the prediction and the features used for the prediction include pickup_community_area, pickup_latitude, and longitude along with the created variables. The feature selection is done based on similar research papers that were done to predict the fare and number of trips for Chicago and New York. Data from 2014 to 2019 is used for training both the models.

Regression was run to identify the best parameters for training the Decision Tree algorithm, and minimum sample leaves are passed as 2 and minimum sample splits are passed as 64, and the result gave 93% of accuracy for the test data with root mean square error of 0.19. Decision trees are prone to overfitting, which makes the results inaccurate and not trustworthy. Decision trees are faster because there is only one tree. The Random Forest model is a strong modeling technique and much more robust than Decision tree. In Random Forest multiple decision trees are used to get the results. Hence this process is slow but more accurate than the decision tree. Random forest model is run with 45 as a number of estimators and 1 minimum sample leaves and 2 minimum sample splits as parameters based on regression results to find the best set of parameters. Random Forest (RF) gave 94.6% accuracy with a root mean square error of 0.173 meaning the results from RF are closer to the actual values of the test data which is 20% of the data from 2014 to 2019 than the results of DT.

As the results looked good for test data, these models are then used to predict the number of trips for 2020. The below figure shows how the results are, and how they differed from the actual values because of Covid-19. Random forest and Decision Tree were used to predict the number of trips for the years 2014 to 2020. It was found that both of these models had similar results. These results were also like the actual data for 2014 to early 2020. Starting from March 2020, the actual results deviated drastically from the predicted with a sharp decline in the number of trips (Fig. 15).

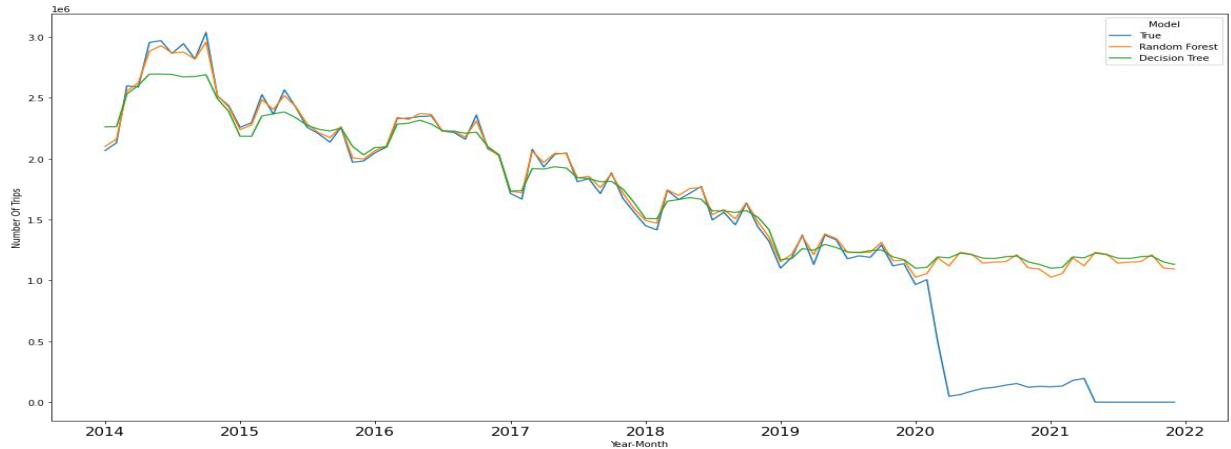

Figure 15: Predicted vs Actual Yearly representation of the number of trips.

*2021 So Far and What to Expect*

Even after a year of the pandemic, the number of trips is still very low in number comparatively, at the same time we can see that there is a considerable increase compared to 2020 and growth is consistent so far (fig.16). There are more than 0.23 Million trips reported in the month of April-2021 which is a 200% increase from April-2020 and a 35% increase from December-2020. Taxi drivers are still waiting for hours to get the passengers. With more people being vaccinated, social-economic activity is slowly going back to normal, and when offices and colleges start to reopen for in-person meetings and classes we may see more trips taking place.

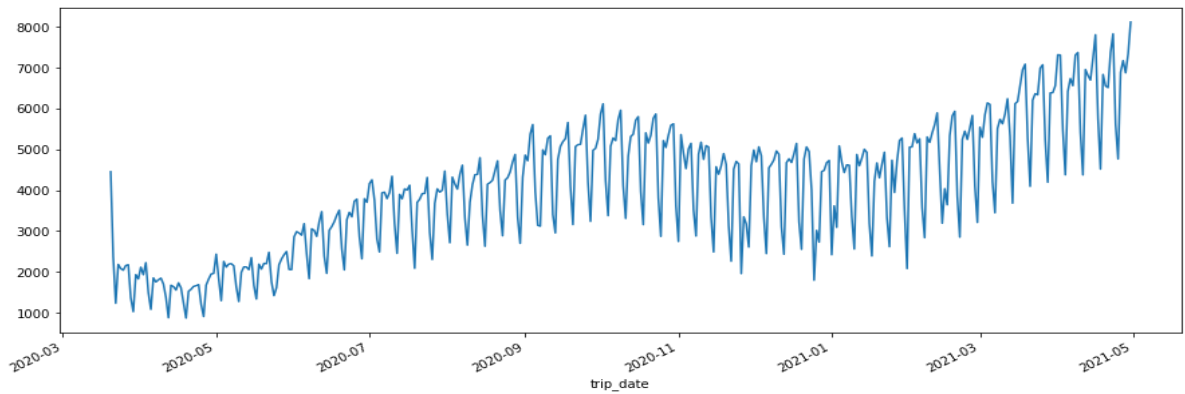

Figure 16: Taxi trips of 2021

**Conclusion**

There was an overall decrease in the number of taxi trips in 2020 due to Covid-19. The taxi industry suffered in many ways. The number of trips has gone down by more than 90%. There was also a reduction in the average tip amount which made it hard for the drivers who were already struggling with a reduction of trips. The number of active taxi licenses also decreased to around 800 from 6000-8000 in previous years. People tended to travel longer distances during the pandemic, but the trip times were shorter which may be due to less traffic from stay-at-home orders. Some of the most popular pick-up and drop-off locations experienced up to a 95% decrease in the number of trips. However, there were some places such as the Far Southeast Side where there was an increase in trips throughout the years and even during the pandemic. Some places also had their number of trips double during the pandemic, but these areas contribute only 5% of the total trips. According to the predictions, the trend of trips in 2020 was expected to be linear. However, due to the pandemic, there was a sharp decrease. It was expected to have an average of 60000 trips per day but instead had an average of 3000 trips. Looking at the data from 2021, the taxi industry seems to be improving slowly because of the increase in social-economic activity.